\documentclass[12pt]{article}
\usepackage{graphicx}
\usepackage{color}
\usepackage{authblk}

\def\hybrid{\topmargin 0pt      \oddsidemargin 0pt
        \headheight 0pt \headsep 0pt
       \voffset-1cm
        \textwidth 6.25in       
       \textheight 9.5in       
        \marginparwidth 0.0in
        \parskip 5pt plus 1pt   \jot = 1.5ex}
\catcode`\@=11
\def\marginnote#1{}

\newcount\hour
\newcount\minute
\newtoks\amorpm
\hour=\time\divide\hour by60
\minute=\time{\multiply\hour by60 \global\advance\minute by-\hour}
\edef\standardtime{{\ifnum\hour<12 \global\amorpm={am}%
        \else\global\amorpm={pm}\advance\hour by-12 \fi
        \ifnum\hour=0 \hour=12 \fi
        \number\hour:\ifnum\minute<10 0\fi\number\minute\the\amorpm}}
\edef\militarytime{\number\hour:\ifnum\minute<10 0\fi\number\minute}

\def\draftlabel#1{{\@bsphack\if@filesw {\let\thepage\relax
   \xdef\@gtempa{\write\@auxout{\string
      \newlabel{#1}{{\@currentlabel}{\thepage}}}}}\@gtempa
   \if@nobreak \ifvmode\nobreak\fi\fi\fi\@esphack}
        \gdef\@eqnlabel{#1}}
\def\@eqnlabel{}
\def\@vacuum{}
\def\draftmarginnote#1{\marginpar{\raggedright\scriptsize\tt#1}}

\def\draftlabel#1{{\@bsphack\if@filesw {\let\thepage\relax
   \xdef\@gtempa{\write\@auxout{\string
      \newlabel{#1}{{\@currentlabel}{\thepage}}}}}\@gtempa
   \if@nobreak \ifvmode\nobreak\fi\fi\fi\@esphack}
        \gdef\@eqnlabel{#1}}
\def\@eqnlabel{}
\def\@vacuum{}
\def\draftmarginnote#1{\marginpar{\raggedright\scriptsize\tt#1}}

\def\draft{\oddsidemargin -.5truein
        \def\@oddfoot{\sl preliminary draft \hfil
        \rm\thepage\hfil\sl\today\quad\militarytime}
        \let\@evenfoot\@oddfoot \overfullrule 3pt
        \let\label=\draftlabel
        \let\marginnote=\draftmarginnote
   \def\@eqnnum{(\theequation)\rlap{\kern\marginparsep\tt\@eqnlabel}%
\global\let\@eqnlabel\@vacuum}  }

\def\numberbysection{\@addtoreset{equation}{section}
        \def\theequation{\thesection.\arabic{equation}}}

\def\underline#1{\relax\ifmmode\@@underline#1\else
        $\@@underline{\hbox{#1}}$\relax\fi}

\def\titlepage{\@restonecolfalse\if@twocolumn\@restonecoltrue\onecolumn
     \else \newpage \fi \thispagestyle{empty}\c@page\z@
        \def\thefootnote{\fnsymbol{footnote}} }

\def\endtitlepage{\if@restonecol\twocolumn \else  \fi
        \def\thefootnote{\arabic{footnote}}
        \setcounter{footnote}{0}}  
\relax

\hybrid

\newfont{\Bbb}{msbm10 scaled 1\@ptsize00}
\newfont{\Bbbb}{msbm7 scaled 1\@ptsize00}

\newcommand{\DDD}{\raise-1pt\hbox{$\mbox{\Bbbb D}$}}

\newcommand{\UUU}{\raise-1pt\hbox{$\mbox{\Bbbb U}$}}

\newcommand{\ZZ}{\mbox{\Bbb Z}}
\newcommand{\z}{\raise-1pt\hbox{$\mbox{\Bbbb Z}$}}

\def\res{\mathop{\hbox{res}}\limits}

\def\beq{\begin{equation}}
\def\eeq{\end{equation}}
\def\p{\partial}

\begin{document}

\begin{titlepage}

\title{Elliptic solutions to the KP hierarchy and elliptic Calogero-Moser model}

\vspace{0.5cm}

\author[1,2]{V.~Prokofev\thanks{vadim.prokofev@phystech.edu }}
\author[2,3,4]{
 A.~Zabrodin\thanks{ zabrodin@itep.ru}}
 \affil[1]{Moscow Institute of Physics and Technology, Dolgoprudny, Institutsky per., 9,
Moscow region, 141700, Russia}
 \affil[2]{
Skolkovo Institute of Science and Technology, 143026 Moscow, Russian Federation
}
\affil[3]{ National Research University Higher School of Economics,
20 Myasnitskaya Ulitsa, Moscow 101000, Russian Federation}
\affil[4]{
Institute of Biochemical Physics, Kosygina str. 4, 119334, Moscow, Russia, Russian
Federation
}

\date{February 2021}
\maketitle


\begin{abstract}

We consider solutions of the KP hierarchy which are elliptic functions
of $x=t_1$. It is known that 
their poles as functions of $t_2$ 
move as particles of the elliptic Calogero-Moser model. 
We extend this correspondence to the level of hierarchies
and find the Hamiltonian $H_k$ of the elliptic Calogero-Moser
model which governs the dynamics of poles with respect to the $k$-th
hierarchical time. The Hamiltonians $H_k$ are obtained as coefficients 
of the expansion of the spectral curve near the marked point in which
the Baker-Akhiezer function has essential singularity.

\end{abstract}

\end{titlepage}

\tableofcontents


\section{Introduction}

The investigation of dynamics of poles of singular solutions to nonlinear
integrable equations was initiated in the seminal paper \cite{AMM77}, where it was shown
that poles of elliptic and rational solutions to the Korteweg-de Vries 
and Boussinesq equations move as particles of the 
integrable many-body Calogero-Moser system \cite{Calogero71,Calogero75,Moser75,OP81}
with some restrictions in the phase space.
As it was proved in \cite{Krichever78,CC77},
this connection becomes most natural for the more general 
Kadomtsev-Petviashvili (KP) equation, in which case there are no restrictions in the phase space 
for the Calogero-Moser dynamics of poles. 

The KP equation is the first member of an infinite hierarchy of consistent 
integrable equations with infinitely many 
independent variables (times) ${\bf t}=\{t_1, t_2, t_3, \ldots \}$ (the KP hierarchy).
In \cite{Shiota94}, Shiota has shown that 
the correspondence between rational solutions to the KP equation and the Calogero-Moser
system with rational potential can be extended to the level of hierarchies: 
the evolution of poles with respect to the higher times $t_k$ 
of the KP hierarchy was shown to be governed by 
the higher Hamiltonians $H_k=\mbox{tr}\, L^k$
of the integrable Calogero-Moser system, where $L$ is the Lax matrix. 
Later this correspondence was generalized to trigonometric
solutions of the KP hierarchy (see \cite{Haine07,Z19a}). 

A natural generalization of rational and trigonometric solutions are elliptic
(double periodic in the complex plane) solutions. Elliptic solutions
to the KP equation 
\beq\label{kp3a}
3u_{t_2t_2}=\Bigl (4u_{t_3}-12uu_x -u_{xxx}\Bigr )_x
\eeq
(where $x=t_1$)
were studied by Krichever in \cite{Krichever80}, where
it was shown that
poles $x_i$ of the elliptic solutions 
\beq\label{int0}
u=-\sum_{i=1}^{N}\wp (x-x_i)+2c
\eeq
as functions of $t_2$ move according to the equations of motion
\beq\label{int1}
\ddot x_i=4\sum_{k\neq i} \wp ' (x_i-x_k)
\eeq
of the Calogero-Moser system of particles with the elliptic
interaction potential $\wp (x_i-x_j)$ ($\wp$ is the Weierstrass $\wp$-function).
Here dot means derivative with respect to the time $t_2$. See also the review \cite{Z19}.
The Calogero-Moser system is Hamiltonian with the Hamiltonian
\beq\label{int2}
H=\sum_i p_i^2 -2\sum_{i<j}\wp (x_i-x_j)
\eeq
and the Poisson brackets $\{x_i, p_k\}=\delta_{ik}$. Note that $\dot x_i=
\p H/\p p_i =2p_i$. It is known \cite{Perelomov} 
that the elliptic Calogero-Moser system is integrable,
i.e., there are $N$ independent integrals of motion $H_k$ in involution. 

The aim of this paper is to establish the precise correspondence between the flows of the
KP hierarchy parametrized by the times $t_m$ 
and the Hamiltonian flows of the hierarchy of the elliptic Calogero-Moser systems. 
In short, the result is as follows. 
Let the function $\lambda (z)$ be determined from the equation of the 
Calogero-Moser spectral curve in the form
\beq\label{int3}
\det_{N\times N} \Bigl ((z+\zeta (\lambda ))I-L(\lambda )\Bigr )=0,
\eeq
where $I$ is the unity matrix, $L(\lambda )$ is the Lax matrix of the Calogero-Moser system
depending on the spectral parameter $\lambda$ and $\zeta (\lambda )$ is the Weierstrass 
$\zeta$-function. We show that the function $\lambda (z)$ expanded as $z\to \infty$ as
\beq\label{int4}
\lambda (z)=-Nz^{-1}+\sum_{m\geq 1}H_m z^{-m-1}
\eeq
is the generating function for the Calogero-Moser Hamiltonians $H_m$ corresponding to the 
flows $t_m$ of the KP hierarchy. We find first few Hamiltonians explicitly. In the rational
and trigonometric limit it is possible to find them for any $m$ in terms of traces of the 
Lax matrix and the result coincides with what was previously known (see \cite{Shiota94,Z19a}).

\section{The KP hierarchy}

The KP hierarchy is an infinite set of evolution equations in the times
${\bf t}=\{t_1, t_2, t_3, \ldots \}$ for functions of a variable $x$. 
In the Lax formulation of the hierarchy, the main object is the 
pseudo-differential operator
\beq\label{kp1}
{\cal L}=\p_x +\sum_{k\geq 1}u_k \p_x^{-k},
\eeq
where the coefficient functions $u_k$ are
functions of $x$ and ${\bf t}$. The equations of the KP hierarchy 
are encoded in the 
Lax equations
\beq\label{kp2}
\p_{t_m}{\cal L}=[{\cal A}_m, {\cal L}], \qquad {\cal A}_m=({\cal L}^m)_+,
\eeq
where $(\ldots )_+$ means taking the purely differential part of a pseudo-differential
operator. In particular, we have $\p_{t_1}{\cal L}=\p_x {\cal L}$, i.e., 
$\p_{t_1}u_k =\p_x u_k$ for all $k\geq 1$. This means that the evolution in $t_1$ is simply
a shift of $x$: $u_k (x, {\bf t})=u_k(x+t_1, t_2, t_3, \ldots )$. 

An equivalent formulation of the KP hierarchy is through the zero curvature
(Za\-kha\-rov-\-Sha\-bat) equations
\beq\label{kp3}
\p_{t_n}{\cal A}_m -\p_{t_m}{\cal A}_n +[{\cal A}_m, {\cal A}_n]=0.
\eeq
The simplest nontrivial equation (\ref{kp3a}) is obtained for $u=u_1$ at $m=2$, $n=3$. 

A common solution to the KP hierarchy is provided by the tau-function
$\tau = \tau (x, {\bf t})$. The coefficient functions $u_k$ of the Lax operator
can be expressed through the tau-function. 
For example,
\beq\label{kp10a}
u_1(x, {\bf t})=u(x, {\bf t})=\p_x^2\log \tau (x, {\bf t}).
\eeq
The whole hierarchy is encoded in the bilinear relation \cite{DJKM83,JM83}
\beq\label{kp7}
\oint_{\infty}e^{(x-x')z +\xi ({\bf t}, z)-\xi ({\bf t}', z)}
(e^{-D(z)}\tau (x, {\bf t}))( e^{D(z)}\tau (x', {\bf t}'))dz =0
\eeq
valid for all $x, x'$, ${\bf t}, {\bf t}'$, where
$$
\xi ({\bf t}, z)=\sum_{k\geq 1}t_k z^k
$$
and $D(z)$ is the differential operator
\beq\label{kp8}
D(z)=\sum_{k\geq 1}\frac{z^{-k}}{k}\, \p_{t_k}.
\eeq
The integration contour is a big circle around infinity separating the singularities
coming from the exponential factor from those coming from the tau-functions.

Let us point out an important corollary of the bilinear relation.
Applying the operator 
$\displaystyle{D'(\mu )=-\sum_{k\geq 1}\mu^{-k-1}\p_{t_k}}$  to (\ref{kp7})
and putting $x=x'$, 
${\bf t}={\bf t}'$
after that, we obtain
$$
-\sum_{k\geq 1}\oint_{\infty}\mu^{-k-1}z^k 
(e^{-D(z)}\tau )\, (e^{D(z)}\tau )\, dz +\oint_{\infty} D'(\mu )
(e^{-D(z)}\tau )\, (e^{D(z)}\tau )\, dz =0
$$
or
$$
\frac{1}{2\pi i}\oint_{\infty}\frac{z}{\mu (z-\mu )}\,
(e^{-D(z)}\tau )\, (e^{D(z)}\tau )\, dz=
D'(\mu )\p_x \tau \, \tau - D'(\mu )\tau \, \p_x \tau .
$$
Taking the residues in the left hand side, we get the equation
\beq\label{kp9}
\frac{(e^{D(\mu )}\tau )\, (e^{-D(\mu )}\tau )}{\tau^2}=
1-D'(\mu )\p_x \log \tau .
\eeq

The zero curvature equations (\ref{kp3}) are compatibility conditions of the 
auxiliary linear problems 
\beq\label{kp10}
\p_{t_k}\psi ={\cal A}_k \psi
\eeq
for the wave function $\psi = \psi (x, {\bf t}, z)$ depending on the spectral 
parameter $z$. In particular, at $k=2$ we have the equation
\beq\label{kp11}
\p_{t_2}\psi =\p_x^2\psi +2u\psi .
\eeq
One can also introduce the adjoint wave function $\psi^*$ satisfying the adjoint equation
(\ref{kp10}):
\beq\label{kp12}
-\p_{t_k}\psi^* ={\cal A}_k^{\dag} \psi^*,
\eeq
where the ${}^{\dag}$-operation is defined as $(f(x)\circ \p_x^n)^{\dag}=(-\p_x)^n \circ f(x)$.
In \cite{DJKM83,JM83} it is shown that the wave functions can be expressed through the
tau-function in the following way:
\beq\label{kp13}
\psi (x, {\bf t}, z)=e^{xz+\xi({\bf t}, z)}\frac{e^{-D(z)}\tau (x, {\bf t})}{\tau (x, {\bf t})},
\eeq
\beq\label{kp13a}
\psi^* (x, {\bf t}, z)=e^{-xz-\xi({\bf t}, z)}
\frac{e^{D(z)}\tau (x, {\bf t})}{\tau (x, {\bf t})}.
\eeq
Note that in terms of the wave functions the equation (\ref{kp9}) can be written 
in the form
\beq\label{kp14}
\p_{t_m}\p_{t_1}\log \tau (x, {\bf t})=\res_{\infty} \Bigl (z^m
\psi (x, {\bf t}, z)\psi^* (x, {\bf t}, z)\Bigr ),
\eeq
where $\res_{\infty}$ is defined as $\res_{\infty}(z^{-n})=\delta_{n1}$.

\section{Elliptic solutions}

The ansatz for the tau-function of elliptic (double-periodic in the complex plane)
solutions to the KP hierarchy is
\beq\label{e1}
\tau = e^{Q(x, {\bf t})}\prod_{i=1}^{N}\sigma (x-x_i({\bf t})),
\eeq
where 
$$
Q(x, {\bf t})=c(x+t_1)^2 +(x+t_1)A(t_2, t_3, \ldots ) +B(t_2, t_3, \ldots )
$$
with a constant $c$, a linear function  
\beq\label{e1b}
A(t_2, t_3, \ldots )=A_0 +\sum_{j\geq 2}a_j t_j
\eeq
and some function 
$B(t_2, t_3, \ldots )$. 
In (\ref{e1})
$$
\sigma (x)=\sigma (x |\, \omega , \omega ')=
x\prod_{s\neq 0}\Bigl (1-\frac{x}{s}\Bigr )\, e^{\frac{x}{s}+\frac{x^2}{2s^2}},
\quad s=2\omega m+2\omega ' m' \quad \mbox{with integer $m, m'$}
$$ 
is the Weierstrass 
$\sigma$-function with quasi-periods $2\omega$, $2\omega '$ such that 
${\rm Im} (\omega '/ \omega )>0$. 
It is connected with the Weierstrass 
$\zeta$- and $\wp$-functions by the formulas $\zeta (x)=\sigma '(x)/\sigma (x)$,
$\wp (x)=-\zeta '(x)=-\p_x^2\log \sigma (x)$.
The monodromy properties of the function $\sigma (x)$ 
are
\beq\label{e1a}
\sigma (x+2\omega )=-e^{2\eta (x+\omega )}\sigma (x), \quad
\sigma (x+2\omega ' )=-e^{2\eta ' (x+\omega ' )}\sigma (x),
\eeq
where the constants $\eta =\zeta (\omega )$, $\eta ' =\zeta (\omega ' )$ are  related by
$\eta \omega ' -\eta ' \omega =\pi i /2$. 
The roots $x_i$ are assumed to be 
all distinct. Correspondingly, 
the function $u=\p_x^2\log \tau$ is an 
elliptic function with double poles at the points $x_i$:
\beq\label{e2}
u=-\sum_{i=1}^{N}\wp (x-x_i)+2c.
\eeq
The poles depend on the times $t_1, t_2, t_3, \ldots$. The dependence on $t_1$ 
is especially simple: since the solution must depend on $x+t_1$, we have 
$\p_{t_1}x_i=-1$.

Let $\Delta (\mu )$ be the difference operator
\beq\label{Delta}
\Delta (\mu )=e^{D(\mu )}+e^{-D(\mu )}-2.
\eeq
Substituting the ansatz (\ref{e1}) into equation (\ref{kp9}), we get:
$$
e^{G+(x+t_1)\Delta (\mu )A}\prod_i 
\frac{\sigma (x-e^{D(\mu )}x_i)\, \sigma (x-e^{-D(\mu )}x_i)}{\sigma^2 (x-x_i)}
$$
$$
=\,
1+2c\mu^{-2} -D'(\mu )A -\sum_k D'(\mu )x_k \, \wp (x-x_k),
$$
where
$$
G=2c\mu^{-2}+\mu^{-1}(e^{D(\mu )}-e^{-D(\mu )})A +\Delta (\mu )B.
$$
The right hand side is an elliptic function of $x$ with periods $2\omega$, $2\omega '$.
Therefore, for the left hand side 
be also an elliptic function of $x$ with the same periods 
the following relations have to be satisfied:
$$
\left \{
\begin{array}{l}
\displaystyle{\exp \Bigl (-2\eta \Delta (\mu )\sum_k x_k +2\omega \Delta (\mu)A\Bigr )=1}
\\ \\
\displaystyle{\exp \Bigl (-2\eta ' \Delta (\mu )\sum_k x_k +2\omega ' \Delta (\mu)A\Bigr )=1}
\end{array}
\right.
$$
from which it follows that
$$
\Delta (\mu)A =2n' \eta -2n\eta ', \quad
\Delta (\mu )\sum_k x_k =2n'\omega -2n\omega ' \quad
\mbox{with integer $n,n'$}.
$$
The right hand sides do not depend on $\mu$.
Expanding the equalities in powers of $\mu$, one sees that the left hand sides are
$O(\mu^{-2})$ as $\mu \to \infty$, therefore,
$n=n'=0$ and we have
\beq\label{e4}
\Delta (\mu)A =0, \quad \Delta (\mu )\sum_k x_k =0.
\eeq
The first equation is satisfied if $A$ is a linear function of times as in (\ref{e1b}).
The second equation means that
\beq\label{e4a}
(1-e^{-D(\mu)})\sum_i x_i = -(1-e^{D(\mu)})\sum_i x_i.
\eeq

Note that the functions (\ref{kp13}), (\ref{kp13a}) with $\tau$ as in (\ref{e1}) 
are {\it double-Bloch functions}, i.e., they satisfy the monodromy properties
$\psi (x+2\omega )=B\psi (x)$, $\psi (x+2\omega ' )=B'\psi (x)$ with some
Bloch multipliers $B$, $B'$. Any non-trivial double-Bloch function (i.e. not an exponential
function) must have poles in $x$ in the fundamental domain. The Bloch multipliers of the function (\ref{kp13}) are
\beq\label{bloch}
B=e^{2\omega (z-\alpha (z)) -2\zeta (\omega )(e^{-D(z)}-1)\sum_i x_i}, \qquad
B'=e^{2\omega  ' (z-\alpha (z)) -2\zeta (\omega ')(e^{-D(z)}-1)\sum_i x_i},
\eeq
where
\beq\label{alpha}
\alpha (z)=2cz^{-1}+\sum_{j\geq 2}\frac{a_j}{j}\, z^{-j}
\eeq
with the constants $a_j$ entering (\ref{e1b}).
Equation (\ref{e4a}) means that the Bloch multipliers of the adjoint wave function
$\psi^*$ are $B^{-1}$ and $B^{' -1}$. 

Let us introduce the 
elementary double-Bloch function $\Phi (x, \lambda )$ defined as
\beq\label{Phi}
\Phi (x, \lambda )=\frac{\sigma (x+\lambda )}{\sigma (\lambda )\sigma (x)}\,
e^{-\zeta (\lambda )x}
\eeq
($\zeta (\lambda )$ is the Weierstrass $\zeta$-function).
The monodromy properties of the function $\Phi$ are
$$
\Phi (x+2\omega , \lambda )=e^{2(\zeta (\omega )\lambda - \zeta (\lambda )\omega )}
\Phi (x, \lambda ),
$$
$$
\Phi (x+2\omega ' , \lambda )=e^{2(\zeta (\omega ' )\lambda - \zeta (\lambda )\omega ' )}
\Phi (x, \lambda ),
$$
so it is indeed a double-Bloch function.
The function $\Phi$
has a simple pole
at $x=0$ with residue 1: 
$$
\Phi (x, \lambda )=\frac{1}{x}+\alpha_1 x +\alpha_2 x^2 +\ldots , \qquad 
x\to 0,
$$
where $\alpha_1=-\frac{1}{2}\, \wp (\lambda )$, $\alpha_2=-\frac{1}{6}\, \wp '(\lambda )$. 
We will often suppress the second argument of $\Phi$ writing simply 
$\Phi (x)=\Phi (x, \lambda )$. 
We will also need the $x$-derivative 
$\Phi '(x, \lambda )=\p_x \Phi (x, \lambda )$.

Equations (\ref{kp13}), (\ref{kp13a}) and (\ref{e1}) imply that the wave functions
$\psi$, $\psi^*$ have simple poles at the points $x_i$. One can expand the wave functions
using the elementary double-Bloch functions as follows: 
\beq\label{e5}
\psi =e^{xk +t_1(k-z) +\xi({\bf t}, z)}\sum_i c_i \Phi (x-x_i, \lambda )
\eeq
\beq\label{e5a}
\psi^* =e^{-xk -t_1(k-z) -\xi({\bf t}, z)}\sum_i c^*_i \Phi (x-x_i, -\lambda )
\eeq
(this is similar to expansion of a rational function
in a linear combination of simple fractions). Here $c_i$, $c^*_i$
are expansion coefficients which do not depend on $x$ and 
$k$ is an additional spectral parameter. Note that the normalization of the functions
(\ref{kp13}), (\ref{kp13a}) implies 
that $c_i$ and $c_i^*$ are $O(\lambda )$ as $\lambda \to 0$.
One can see that (\ref{e5}) is 
a double-Bloch function with Bloch multipliers
\beq\label{e6}
B=e^{2\omega (k-\zeta (\lambda )) + 2\zeta (\omega )\lambda }, \qquad
B '=e^{2\omega ' (k-\zeta (\lambda )) + 2\zeta (\omega ' )\lambda }
\eeq
and (\ref{e5a}) has Bloch multipliers $B^{-1}$ and $B^{' -1}$. These Bloch multipliers
should coincide with (\ref{bloch}). 

Therefore, comparing (\ref{bloch}) with (\ref{e6}), we get
$$
2\omega (k-\zeta (\lambda )-z+\alpha (z))+
2\zeta (\omega )\Bigl (\lambda +(e^{-D(z)}-1)\sum_i x_i\Bigr )=
2\pi i n,
$$
$$
2\omega ' (k-\zeta (\lambda )-z+\alpha (z))+
2\zeta (\omega ' )\Bigl (\lambda +(e^{-D(z)}-1)\sum_i x_i\Bigr )=
2\pi i n'
$$
with some integer $n, n'$. Regarding these equations as a linear system, we obtain the solution
$$
k-z+\alpha (z)-\zeta (\lambda)=2n'\zeta (\omega)-2n\zeta (\omega '),
$$
$$
\lambda +(e^{-D(z)}-1)\sum_i x_i=2n\omega ' -2n'\omega .
$$
Shifting $\lambda$ by a suitable vector of the lattice spanned by $2\omega$, $2\omega '$,
one gets zeros in the right hand sides of these equalities, so we can write
\beq\label{e7}
\begin{array}{l}
k=z-\alpha (z)+\zeta (\lambda ),
\\ \\
\displaystyle{\lambda = (1-e^{-D(z)}) \sum_i x_i}.
\end{array}
\eeq
These two equations for three variables $k, z, \lambda$ determine the spectral curve. 
Below we will obtain another description of the spectral curve as the spectral curve
of the Calogero-Moser system (given by the characteristic polynomial of the Lax matrix
$L(\lambda )$ for the Calogero-Moser system). It appears in the form $R(k,\lambda)=0$, where
$R(k,\lambda)$ is a polynomial in $k$ whose coefficients are elliptic functions of
$\lambda$ (see below in section \ref{section:spectral}). 
These coefficients are integrals of motion in involution. The spectral curve
in the form $R(k,\lambda)=0$ appears if one excludes $z$ from the equations (\ref{e7}).
Equivalently, one can represent the spectral curve as a relation connecting two variables
$z$ and $\lambda$:
\beq\label{e7a}
R(z-\alpha (z)+\zeta (\lambda ), \lambda )=0.
\eeq

Let us write the second equation in (\ref{e7}) as the expansion in powers of $z$:
\beq\label{h2}
\lambda = -\sum_{m\geq 1}z^{-m}\hat h_m X, \qquad X:=\sum_i x_i,
\eeq
where $\hat h_k$ are differential operators of the form
\beq\label{h1}
\hat h_m = -\frac{1}{m}\, \p_{t_m}+\,\,\, \mbox{higher order operators
in $\p_{t_1}, \p_{t_2}, \ldots , \p_{t_{m-1}}$}.
\eeq
For example, the first few are
$$
\hat h_1 =-\p_{t_1}, \quad 
\hat h_2 =\frac{1}{2}\, (\p_{t_1}^2 -\p_{t_2}),
\hat h_3 =\frac{1}{6}\, (-\p_{t_1}^3 +3\p_{t_1}\p_{t_2}-2\p_{t_3}).
$$
As is explained above, the coefficients in the expansion (\ref{h2}) are integrals of
motion, i.e., $\p_{t_j}\hat h_m X=0$ for all $j,m$. It then follows from 
the equation $\p_{t_1}x_i=-1$ and from the explicit form of the operators $\hat h_m$ that
$\p_{t_j}\p_{t_2}X=0$ and $\p_{t_j}\p_{t_3}X=0$. A simple inductive argument then shows that
$\p_{t_j}\p_{t_m}X=0$ for all $j,m$. This means that
$-\hat h_m X=\frac{1}{m}\, \p_{t_m}X$ and $X$ is a linear function of the times:
\beq\label{h3}
X=\sum_i x_i =X_0 -Nt_1 +\sum_{m\geq 2}V_m t_m
\eeq
with some constants $V_m$ (velocities of the ``center of masses'' of the points $x_i$
multiplied by $N$). Therefore, the second equation in (\ref{e7}) can be written as
\beq\label{e8}
\lambda = D(z)\sum_i x_i=-Nz^{-1}+\sum_{j\geq 2}\frac{z^{-j}}{j}\, V_j.
\eeq
In what follows we will show that $H_m=-\frac{1}{m+1}\, V_{m+1}$ are Hamiltonians
for the dynamics of the poles in $t_m$, with $H_2$ being the standard Calogero-Moser
Hamiltonian.

\section{Dynamics of poles with respect to $t_2$}

The coefficient $u$ in the linear problem (\ref{kp11})
\beq\label{kp11a}
\p_{t_2}\psi -\p_x^2\psi -2u\psi =0 
\eeq
is an elliptic function of $x$ of the form (\ref{e2}). Therefore, one can find solutions
which are double-Bloch functions of the form (\ref{e5}).

The next procedure is standard after the work \cite{Krichever80}. We substitute
$u$ in the form (\ref{e2}) and $\psi$ in the form (\ref{e5}) into the left hand side
of (\ref{kp11a}) and cancel the poles at the points $x=x_i$. 
The highest poles are of third order but it is easy to see that they cancel identically.
It is a matter of direct calculation to see that the conditions of cancellation of
second and first order poles have the form
\beq\label{ell5}
c_i\dot x_i=-2kc_i -2\sum_{j\neq i}c_j \Phi (x_i-x_j),
\eeq
\beq\label{ell6}
\dot c_i=(k^2\! -\! z^2+4c-2\alpha_1) c_i -2\sum_{j\neq i}c_j \Phi '(x_i-x_j)-2c_i \sum_{j\neq i}
\wp (x_i-x_j),
\eeq
where dot means the $t_2$-derivative.
Introducing $N\! \times \! N$ matrices
\beq\label{ell7}
L_{ij}=-\frac{1}{2}\, \delta_{ij}\dot x_i -(1-\delta_{ij})\Phi (x_i-x_j),
\eeq
\beq\label{ell8}
M_{ij}=\delta_{ij}(k^2\! -\! z^2+\wp (\lambda )+4c)-2\delta_{ij}
\sum_{k\neq i}\wp (x_i-x_k) -2(1-\delta_{ij})\Phi ' (x_i-x_j),
\eeq
we can write the above conditions as a system of linear equations for the vector
${\bf c}=(c_1, \ldots , c_N)^T$:
\beq\label{ell9}
\left \{ \begin{array}{l}
L(\lambda){\bf c}=k{\bf c}
\\ \\
\dot {\bf c}=M(\lambda) {\bf c}.
\end{array} \right.
\eeq
Differentiating the first equation in (\ref{ell9}) with respect to $t_2$, we arrive at
the compatibility condition of the linear problems (\ref{ell9}):
\beq\label{ell12}
\Bigl (\dot L+[L,M]\Bigr ) {\bf c} =0.
\eeq
The Lax equation $\dot L+[L,M]=0$ is equivalent to the equations of motion 
of the elliptic Calogero-Moser system (see
\cite{Z19} for the detailed calculation). 
Our matrix $M$ differs from the standard one by the term $\delta_{ij}(k^2\! -\! z^2)$ but
it does not affect the compatibility condition.
It follows from the Lax representation that the time evolution is an isospectral
transformation of the Lax matrix $L$, so all traces $\mbox{tr}\, L^m$ and the 
characteristic polynomial $\det (L-kI)$, where $I$ is the unity matrix, 
are integrals of motion.
Note that the Lax matrix is written in terms of the momenta $p_i$ as follows:
\beq\label{e9}
L_{ij}=-\delta_{ij}p_i -(1-\delta_{ij})\Phi (x_i-x_j).
\eeq

A similar calculation shows that the adjoint linear problem for the function (\ref{e5a})
leads to the equations
\beq\label{ell9a}
\left \{ \begin{array}{l}
{\bf c}^{*T}L(\lambda)=k{\bf c}^{*T}
\\ \\
\dot {\bf c}^{*T}=-{\bf c}^{*T}M(\lambda) 
\end{array} \right.
\eeq
with the compatibility condition ${\bf c}^{*T}\Bigl (\dot L+[L,M]\Bigr )  =0$.

\section{The spectral curve}

\label{section:spectral}

The first of the equations (\ref{ell9}) determines a connection between
the spectral parameters $k, \lambda$ which is the equation of the spectral curve:
\beq\label{spec1}
R(k, \lambda )=\det \Bigl (kI-L(\lambda )\Bigr )=0.
\eeq
As it was already mentioned, the spectral curve 
is an integral of motion.
The matrix
$L=L(\lambda )$, which has an essential singularity at $\lambda =0$, can be 
represented in the form $L=V\tilde L V^{-1}$, where matrix elements of 
$\tilde L$ do not have 
essential singularities and $V$ is the diagonal matrix $V_{ij}=\delta_{ij}
e^{-\zeta (\lambda )x_i}$. Therefore, 
$$
R(k, \lambda )=\sum_{m=0}^{N}R_m(\lambda )k^m,
$$
where the coefficients $R_m(\lambda )$ are elliptic functions of $\lambda$ with poles
at $\lambda =0$.
The functions 
$R_m (\lambda )$ can be represented as linear combinations of the $\wp$-function and
its derivatives. Coefficients
of this expansion are integrals of motion. Fixing values of these integrals, we obtain
via the equation $R(k, \lambda )=0$ an algebraic curve $\Gamma$ which is a 
$N$-sheet covering of the initial elliptic curve ${\cal E}$ realized as a factor
of the complex plane with respect to the lattice generated by $2\omega$, $2\omega '$.

\noindent
{\it Example} ($N=2$):
$$
\det_{2\times 2}\Bigl (kI-L(\lambda )\Bigr )=k^2 +k(p_1 +p_2)
+p_1p_2 +\wp (x_1-x_2)-\wp (\lambda ).
$$
{\it Example} ($N=3$):
$$
\det_{3\times 3}\Bigl (kI-L(\lambda )\Bigr )=k^3+k^2 (p_1+p_2+p_3)
$$
$$
+k \Bigl (p_1p_2 +p_1p_3 +p_2p_3 +
\wp (x_{12})+\wp (x_{13})+\wp (x_{23})-3\wp (\lambda )\Bigr )
$$
$$
+p_1p_2p_3 +p_1\wp (x_{23})+p_2\wp (x_{13})+
p_3\wp (x_{12})-\wp (\lambda )(p_1+p_2+p_3)-\wp '(\lambda ),
$$
where $x_{ik}=x_i-x_k$.  

In a neighborhood of $\lambda =0$   
the matrix $\tilde L$
can be written as
$$
\tilde L=-\lambda ^{-1}(E-I)+O(1),
$$ 
where
$E$ is the rank $1$ matrix with matrix elements $E_{ij}=1$ for all $i,j =1, \ldots , N$.
The matrix $E$ has eigenvalue $0$ with multiplicity $N-1$ and another eigenvalue 
equal to $N$. Therefore, we can write $R(k, \lambda )$ in the form
\beq\label{spec2}
\begin{array}{lll}
R(k, \lambda )&=&
\det \Bigl (kI+\lambda ^{-1}(E-I)+O(1)\Bigr )
\\ && \\
&=&\displaystyle{ \Bigl (k+(N\! -\! 1)\lambda ^{-1}-f_N(\lambda )\Bigr )\prod_{i=1}^{N-1}
(k-\lambda ^{-1}-f_i(\lambda ))},
\end{array}
\eeq
where $f_i$ are regular functions of $\lambda$ at $\lambda =0$: $f_i(\lambda )=O(1)$ as
$\lambda \to 0$. 
This means that the function $k$
has simple poles on all sheets at the points $P_j$ ($j=1, \ldots , N$) of the curve
$\Gamma$ located 
above $\lambda =0$. Its expansion in the
local parameter $\lambda$ on the sheets near these points is given by the multipliers
in the right hand side of (\ref{spec2}):
\beq\label{spec2a}
\begin{array}{l}
k=\, \lambda^{-1}+f_j(\lambda ) \quad \mbox{near $P_j$}, \quad j=1, \ldots , N-1,
\\ \\
k=-(N\! -\! 1)\lambda ^{-1}+f_{N}(\lambda ) \quad \mbox{near $P_{N}$}.
\end{array}
\eeq
The $N$-th sheet is distinguished, as it can be seen 
from (\ref{spec2a}). As in \cite{Krichever80}, we call it the upper sheet.
Note that equations (\ref{e7}), (\ref{e8}) imply
$$
k(\lambda )=-\frac{N-1}{\lambda}+O(1) \quad \mbox{as $\lambda \to 0$},
$$
so the expansion (\ref{e8}) is the expansion of $\lambda (z)$
on the upper sheet of the spectral curve in a neighborhood of the point $P_N$. 

\section{Dynamics in higher times}

Our basic tool is equation (\ref{kp14}). Substituting $\tau (x, {\bf t})$ in the form
(\ref{e1}) and $\psi$, $\psi^*$ in the form (\ref{e5}), (\ref{e5a}) in it, we have:
\beq\label{ht1}
\sum_i \p_{t_m}x_i \wp (x-x_i)+C(t_2, t_3, \ldots )=
\res_{\infty} \left (z^m \sum_{i,j}c_ic_j^* \Phi (x-x_i, \lambda )\Phi (x-x_j, -\lambda )
\right ).
\eeq
Equating the coefficients in front of the second order poles at $x=x_i$, we obtain
\beq\label{ht2}
\p_{t_m}x_i =\res_{\infty}\Bigl (z^m c_i^* c_i\Bigr )=
\res_{\infty}\Bigl (z^m {\bf c}^{*T}E_i {\bf c}\Bigr ),
\eeq
where $E_i$ is the diagonal matrix with 1 at the place $ii$ and zeros otherwise.
At $m=1$ this reads $\res_{\infty}\Bigl (z {\bf c}^{*T}E_i {\bf c}\Bigr )=-1$ or
$$
\res_{\infty}\Bigl (z \, {\bf c}^{*T}{\bf c}\Bigr )=-N.
$$
Summing the equations (\ref{ht2}) over $i$, we get
$$
\res_{\infty}\Bigl (z^m \, {\bf c}^{*T}{\bf c}\Bigr )=\p_{t_m}\sum_i x_i.
$$
It then follows from these equations that
\beq\label{ht3}
({\bf c}^{*T}{\bf c})= -N/z^2 +\sum_m z^{-m-1}\p_{t_m}\sum_i x_i =-\lambda '(z).
\eeq
The absence of terms with non-negative powers of $z$ in the right hand side 
(which would not change the residue) follows from the
above mentioned fact that ${\bf c}$ and ${\bf c}^*$ are $O(\lambda )=O(z^{-1})$ as $z\to \infty$.
The last equality in (\ref{ht3}) 
follows from (\ref{e8}). Equation (\ref{ht3}) is an important
non-trivial relation which will allow us to identify the Hamiltonians for the higher flows
$t_m$.

Now let us note that according to (\ref{e9}) $E_i=-\p_{p_i}L$. Therefore, we can 
continue the chain of equalities (\ref{ht2}) as follows:
$$
\p_{t_m}x_i =
\res_{\infty}\Bigl (z^m {\bf c}^{*T}E_i {\bf c}\Bigr )=
-\res_{\infty}\Bigl (z^m {\bf c}^{*T}\p_{p_i}L {\bf c}\Bigr )
$$
$$
=-\p_{p_i}\res_{\infty}\Bigl (z^m {\bf c}^{*T}L {\bf c}\Bigr )
+\res_{\infty}\Bigl (z^m \p_{p_i}{\bf c}^{*T}L {\bf c}\Bigr )+
\res_{\infty}\Bigl (z^m {\bf c}^{*T}L \p_{p_i}{\bf c}\Bigr )
$$
$$
=-\p_{p_i}\res_{\infty}\Bigl (z^m {\bf c}^{*T}L {\bf c}\Bigr )
+\res_{\infty}\Bigl (z^m \p_{p_i}{\bf c}^{*T}k {\bf c}\Bigr )+
\res_{\infty}\Bigl (z^m {\bf c}^{*T}k \p_{p_i}{\bf c}\Bigr )
$$
$$
=-\p_{p_i}\res_{\infty}\Bigl (z^m {\bf c}^{*T}L {\bf c}\Bigr )
+\p_{p_i}\res_{\infty}\Bigl (z^m {\bf c}^{*T}k {\bf c}\Bigr )-
\res_{\infty}\Bigl (z^m \p_{p_i}k\, {\bf c}^{*T} {\bf c}\Bigr )
$$
$$
=-\p_{p_i}\res_{\infty}\Bigl (z^m {\bf c}^{*T}(L-kI) {\bf c}\Bigr )-
\res_{\infty}\Bigl (z^m \p_{p_i}k\, {\bf c}^{*T} {\bf c}\Bigr )
$$
$$
=\res_{\infty}\Bigl (z^m \lambda '(z)\p_{p_i}k \Bigr ).
$$
Here $\p_{p_i}k=\p_{p_i}k(\lambda , {\bf I})\Bigr |_{\lambda ={\rm const}}$, where
${\bf I}$ is the full set of integrals of motion. From (\ref{e7}) we see that
\beq\label{ht4}
\p_{p_i}k=(1-\alpha '(z))\, \p_{p_i}z(\lambda , {\bf I})\Bigr |_{\lambda ={\rm const}}.
\eeq
We consider $z$ as an independent variable, so we can write
$$
0=\frac{dz}{dp_i}=\p_{p_i}z\Bigr |_{\lambda ={\rm const}}+\p_{\lambda}z
\Bigr |_{\, {\bf I} ={\rm const}}\p_{p_i}\lambda
$$
or
$$
\p_{p_i}z=-\frac{\p_{p_i} \lambda}{\lambda '(z)}.
$$
Therefore, we have the first set of the Hamiltonian equations
\beq\label{ht5}
\p_{t_m}x_i =-\res_{\infty}\Bigl (z^m (1-\alpha '(z))\p_{p_i} \lambda \Bigr )=
\p_{p_i}{\cal H}_m,
\eeq
where the Hamiltonian
\beq\label{ht5b}
{\cal H}_m =H^{(\alpha )}_m +2cH^{(\alpha )}_{m-2}+\sum_{j=2}^{m-1}
a_j H_{m-j-1}^{(\alpha )}
\eeq
is the linear combination of the Hamiltonians
\beq\label{ht6}
H^{(\alpha )}_m=-\res_{\infty}\Bigl (z^m\lambda (z) \Bigr ) 
\eeq
with constant coefficients. The latter implicitly depend on $\alpha (z)$ through the 
parametrization of the spectral curve (\ref{e7a}). 

In their turn, the Hamiltonians
$H_m^{(\alpha )}$ are linear combinations of the basic Hamiltonians $H_m$ defined at
$\alpha (z)=0$ by
\beq\label{ht8}
H_m =-\res_{\infty}\Bigl (z^m \lambda_0(z)\Bigr ),
\eeq
where $\lambda_0(z)$ is defined through the equation of the spectral curve
\beq\label{e7b}
R(z+\zeta (\lambda_0 ), \lambda_0 )=\det \Bigl ((z+\zeta (\lambda_0))I-L(\lambda_0)\Bigr )=0.
\eeq
Then 
$$
\lambda(z)=\lambda_0(z-\alpha (z))=\lambda_0(z)-\alpha (z)\lambda_0'(z)+
\frac{1}{2}\, \alpha^2(z)\lambda_0''(z)+\ldots 
$$
and so we see that the Hamiltonians (\ref{ht6}) are indeed linear combinations of the
$H_m$'s with constant coefficients.

The remaining set of Hamiltonian equations can be obtained by differentiating
(\ref{ht2}) with respect to $t_2$ and using (\ref{ell9}), (\ref{ell9a}):
$$
2\p_{t_m}p_i=\p_{t_m}\dot x_i=\res_{\infty}\Bigl (z^m {\bf \dot c}^{*T}E_i {\bf c}\Bigr )
+\res_{\infty}\Bigl (z^m {\bf  c}^{*T}E_i {\bf \dot c}\Bigr )=
\res_{\infty}\Bigl (z^m {\bf c}^{*T}[E_i, M] {\bf c}\Bigr )
$$
Now, it is a matter of direct verification to see that
\beq\label{ht7}
[E_i, M]=2\p_{x_i}L.
\eeq
Therefore, we can write
$$
\p_{t_m}p_i=\res_{\infty}\Bigl (z^m {\bf c}^{*T}\p_{x_i}L {\bf c}\Bigr ).
$$
Repeating the transformations presented above in detail, we have:
$$
\p_{t_m}p_i=\res_{\infty}\Bigl (z^m {\bf  c}^{*T}{\bf c}\p_{x_i}k\Bigr )=
-\res_{\infty}\Bigl (z^m \lambda '(z)\p_{x_i}k\Bigr )
$$
The same argument as above shows that
\beq\label{ht4a}
\p_{x_i}k=(1-\alpha '(z))\, \p_{x_i}z(\lambda , {\bf I})\Bigr |_{\lambda ={\rm const}}
\eeq
and
$$
\p_{x_i}z=-\frac{\p_{x_i} \lambda}{\lambda '(z)}.
$$
Therefore, we obtain the second set of Hamiltonian equations for the dynamics of poles:
\beq\label{ht5a}
\p_{t_m}p_i =\res_{\infty}\Bigl (z^m (1-\alpha '(z))\p_{x_i} \lambda \Bigr )=
-\p_{x_i}{\cal H}_m.
\eeq

Let us find ${\cal H}_m$ explicitly in terms of $H_m$ for the case when $a_j=0$, $c\neq 0$.
In this case $\alpha (z)=2cz^{-1}$ and we have
\beq\label{ht9}
\lambda (z)=\lambda_0(z)-\sum_{j,n\geq 1} (2c)^n \left (
\begin{array}{c}n\! +\! j\! -\! 1\\ n\end{array}\right )H_{j-1}z^{-j-2n},
\eeq
${\cal H}_m=H_m^{(\alpha )}+2c H_{m-2}^{(\alpha )}$ and from (\ref{ht9}) we see that
$$
H_m^{(\alpha )}=H_m +\sum_{j=1}^{[m/2]} (2c)^j 
\left (\begin{array}{c}m\! -\! j\!\\ j\end{array}\right )H_{m-2j}.
$$
Therefore,
\beq\label{ht10}
{\cal H}_m=H_m +\sum_{j=1}^{[m/2]} (2c)^j\left [
\left (\begin{array}{c}m\! -\! j\!\\ j\end{array}\right )+
\left (\begin{array}{c}m\! -\! j\! +\! 1\\ j-1\end{array}\right )\right ]H_{m-2j}.
\eeq
In particular, ${\cal H}_3=H_3+6cH_1$ which agrees with the result of the paper \cite{Z19}.

\section{Calculation of the Hamiltonians}

In order to find the Hamiltonians explicitly, we use the description
of the spectral curve given in the paper \cite{IM96}:
\beq\label{ham1}
\sum_{j=0}^N I_j T_{N-j}(k|\lambda )=0,
\eeq
where $T_{j}(k|\lambda )$ are polynomials in $k$ of degree $N$ such that
\beq\label{ham2}
\p_k T_{n}(k|\lambda )=nT_{n-1}(k|\lambda )
\eeq
and $I_j$ are integrals of motion. The first few are
\beq\label{ham3}
\begin{array}{l}
\displaystyle{I_0=1,}
\\ \\
\displaystyle{I_1=\sum_i p_i,}
\\ \\
\displaystyle{I_2=\sum {}^{'} \left (\frac{1}{2!}\, p_ip_j +\frac{1}{2!}\, \wp (x_{ij})\right ),}
\\ \\
\displaystyle{I_3=\sum {}^{'} \left (\frac{1}{3!}\, p_ip_jp_k +\frac{1}{2!}\, p_i
\wp (x_{jk})\right ),}
\\ \\
\displaystyle{I_4=\sum {}^{'} \left (\frac{1}{4!}\, p_ip_jp_kp_l +\frac{1}{2! \cdot 2!}\, p_ip_j
\wp (x_{kl})+ \frac{1}{2 \cdot (2!)^2}\, 
\wp (x_{ij})\wp (x_{kl})\right ),}
\\ \\
\displaystyle{I_5=\sum {}^{'} \left (\frac{1}{5!}\, p_ip_jp_kp_lp_r +\frac{1}{2! \cdot 3!}\, p_ip_jp_r
\wp (x_{kl})+ \frac{1}{2 \cdot (2!)^2}\, p_r
\wp (x_{ij})\wp (x_{kl})\right ),}
\end{array}
\eeq
where $\sum {}^{'}$ means summation over distinct indices. 
Recalling the equation of the spectral curve in terms of $z$ and $\lambda$, let us also
introduce $S_n(z|\lambda )=T_n(z+\zeta (\lambda )|\lambda)$, then
\beq\label{ham4}
\p_z S_{n}(z|\lambda )=nS_{n-1}(z|\lambda ).
\eeq
For example,
$$
T_5(k|\lambda )=k^5 -10\wp (\lambda )k^3 -10 \wp '(\lambda )k^2 -
5(\wp ''(\lambda )-3\wp ^2(\lambda ))k -2\wp '(\lambda )\wp (\lambda ).
$$

We have
$$
\zeta (\lambda )=\frac{1}{\lambda}-\frac{g_2\lambda^3}{2^2\cdot 3\cdot 5}+O(z^{-5}),
$$
where
$$
g_2=60\sum_{s\neq 0}\frac{1}{s^4}, \qquad s=2m\omega +2m'\omega ', \quad m,m'\in \ZZ .
$$
Expanding $S_5(z|\lambda )$ in $z$ using the above formula
for $T_5$, we get:
$$
S_5(z|\lambda )=z^5 +\frac{5z^4}{\lambda }-\frac{g_2}{2}\Bigl (
\frac{1}{\lambda}+5z +5z^2\lambda +\frac{10}{3}\, z^3\lambda^2 +\frac{1}{6}\, z^4\lambda^3 
\Bigr )+O(z^{-1}).
$$
Note that if we introduce the gradation such that $\mbox{deg}\, z=1$, $\mbox{deg}\, \lambda =-1$,
then $\mbox{deg}\, g_2=4$, $\mbox{deg}\, S_n=n$. Note also that in the rational limit
$g_2=0$ and the equation of the spectral curve becomes linear in $\lambda^{-1}$ (see below
in the next section). This can be only in the case if
$S_{n}(z|\lambda )=z^n -nz^{n-1}\lambda^{-1}$ in the rational limit (the coefficient is found
from the condition (\ref{ham4})). 

In the non-degenerate case
we have
\beq\label{ham5}
S_{n}(z|\lambda )=z^n -\frac{nz^{n-1}}{\lambda}+g_2 O(z^{n-4}) 
\eeq
or
\beq\label{ham6}
S_{n}(z|\lambda )=z^n -\frac{nz^{n-1}}{\lambda}+g_2A_n z^{n-4}+g_2B_n I_1 z^{n-5}+O(z^{n-6}),
\eeq
where $A_n$ and $B_n$ are some constant coefficients. ($I_1$ comes from the 
expansion $\lambda=-Nz^{-1}-\displaystyle{\frac{I_1}{2}}z^{-2}+O(z^{-3})$.)  
Therefore, we can write
\beq\label{ham7}
\begin{array}{l}
\displaystyle{S_{N}(z|\lambda )=z^N -
\frac{Nz^{N-1}}{\lambda}+g_2(A_N z^{N-4}+B_N I_1 z^{N-5})+O(z^{N-6}),}
\\ \\
\displaystyle{S_{N-1}(z|\lambda )=z^{N-1} -
\frac{(N-1)z^{N-2}}{\lambda}+g_2A_{N-1} z^{N-5}+O(z^{N-6}),}
\\ \\
\displaystyle{S_{N-j}(z|\lambda )=z^{N-j} -
\frac{(N-j)z^{N-j-1}}{\lambda}+O(z^{N-6}), \quad j=2,3,4,5.}
\end{array}
\eeq
and the equation of the spectral curve acquires the form
$$
z^{N}+\sum\limits_{i=1}^5I_iz^{N-i}+g_2A_{N}z^{N-4}
+g_2(A_{N-1} + B_N)I_1z^{N-5}+O(z^{N-6})
$$
$$
=-\frac{1}{\lambda}\Bigl (Nz^{N-1} +\sum\limits_{i=1}^5(N-i)I_iz^{N-i}
\Bigr ).
$$
Expressing $\lambda$ as a function of $z$ from here, we have:
\beq\label{ham8}
\begin{array}{l}
H_1=-I_1,
\\ \\
H_2=I_1^2-2I_2,
\\ \\
H_3=-I_1^3 +3I_1I_2 -3I_3,
\\ \\
H_4=I_1^4-4I_1^2I_2 +2I_2^2 +4I_1I_3 -4I_4 +\mbox{const},
\\ \\
H_5=-I_1^5 +5I_1^3 I_2 -5I_1^2I_3 +5I_2I_3 -5I_1I_2^2 +5I_1I_4 -5I_5
+g_2 K I_1 ,
\end{array}
\eeq
where $K=(N+1)A_N-N (A_{N-1}+B_N)$,
or, explicitly,
\beq\label{ham9}
\begin{array}{l}
\displaystyle{H_1=-\sum_i p_i},
\\ \\
\displaystyle{H_2=\sum_i p_i^2 -\sum_{i\neq j}\wp (x_{ij}),}
\\ \\
\displaystyle{H_3=-\sum_i p_i^3 +3\sum_{i\neq j}p_i\wp (x_{ij}),}
\\ \\
\displaystyle{H_4=\sum_i p_i^4 \!-\!2\sum_{i\neq j}p_ip_j\wp (x_{ij})-4
\sum_{i\neq j}p_i^2 \wp (x_{ij})}
\\ \\
\displaystyle{\phantom{aaaaaaaaaaa}+\sum_{i\neq j}\wp^2 (x_{ij})+2\sum {}^{'}\wp (x_{ij})\wp (x_{jk})+\mbox{const},}
\\ \\
\displaystyle{H_5=-\sum_i p_i^5 +5\sum_{i\neq j}(p_i^3+p_i^2p_j)\wp (x_{ij})-5
\sum_{i\neq j}p_i \wp ^2 (x_{ij})}
\\ \\
\displaystyle{\phantom{aaaaaaaaaaa}-5
\sum {}^{'}p_i \wp (x_{ij})\wp (x_{ik})-5\sum {}^{'}p_i\wp (x_{ij})\wp (x_{jk})
+\mbox{const}\cdot \sum_i p_i.}
\end{array}
\eeq
These are indeed the Hamiltonians of the elliptic Calogero-Moser model. 
It is easy to see that they satisfy the property
\beq\label{ham11}
H_{m-1}=-\frac{1}{m}\sum_i \p_{p_i}H_m.
\eeq
Indeed, we have
$$
\lambda (z)=-Nz^{-1}+\sum_{m\geq 2}\frac{z^{-m}}{m}\, V_m =
-Nz^{-1}+\sum_{m\geq 1}z^{-m-1}H_m
$$
so
$$
V_m=\p_{t_m}\sum_i x_i =\sum_i \p_{p_i} H_m =-mH_{m-1}.
$$

One can see that the higher Hamiltonians will consist from the principal part
and other terms as follows:
\beq\label{ham10}
H_n=(-1)^n \sum_{|\mu |=n} C_{\mu}^n I_{\mu}+g_2\sum_{|\nu |=n-4} B_{\nu}^n I_{\nu}+\ldots ,
\eeq
where the first sum is taken over Young diagrams $\mu$ of $n=|\mu |$ boxes, 
$I_{\mu}=I_{\mu_1}I_{\mu_2}\ldots I_{\ell (\mu )}$, where $\ell (\mu )$ is the number of
non-empty rows of the diagram $\mu$ and $C_{\mu}^n$ is the matrix of the transition from the
basis of elementary symmetric polynomials to the basis of power sums.

\section{Rational and trigonometric limits}

In the rational limit $\omega_1, \omega_2 \to \infty$, $\sigma (\lambda )=\lambda$,
$\Phi (x, \lambda )=(x^{-1}+\lambda^{-1})e^{-x/\lambda}$ and the equation of the spectral
curve becomes
\beq\label{rat1}
\det \Bigl (L_{\rm rat}-(E-I)\lambda^{-1}-(z+\lambda^{-1})I\Bigr )=0,
\eeq
where
\beq\label{rat2}
(L_{\rm rat})_{ij}=-\delta_{ij}p_i -\frac{1-\delta_{ij}}{x_i-x_j}
\eeq
is the Lax matrix of the rational Calogero-Moser model. Rewriting the equation of
the spectral curve in the form
$$
\det \left (I-E\frac{\lambda^{-1}}{L_{\rm rat}-zI}\right )=0
$$
and using the property $\det (I+Y)=1+\mbox{tr}\, Y$ for any matrix $Y$ of rank 1, we get
\beq\label{rat3}
\lambda =-\mbox{tr}\Bigl (E\frac{1}{zI-L_{\rm rat}}\Bigr )=-\sum_{n\geq 0}
z^{-n-1}\, \mbox{tr}\, L_{\rm rat}^n,
\eeq
where we use the well known property $\mbox{tr}(EL_{\rm rat}^n)=\mbox{tr}(L_{\rm rat}^n)$. So the
Hamiltonians are $H_m=\mbox{tr} L_{\rm rat}^m$ which agrees with Shiota's result \cite{Shiota94}. 

The trigonometric limit is more tricky. Let $\pi i/\gamma$ be period of the 
trigonometric (or hyperbolic) functions (the second period tends to infinity).
The Weierstrass functions in this limit become
$$
\sigma (x)=\gamma^{-1}e^{-\frac{1}{6}\, \gamma^2x^2}\sinh (\gamma x),
\quad
\zeta (x)=\gamma \coth (\gamma x)-\frac{1}{3}\, \gamma^2 x.
$$
The tau-function for trigonometric solutions is
\beq\label{trig1}
\tau (x, {\bf t})=\prod_{i=1}^N \Bigl (e^{2\gamma x}-e^{2\gamma x_i({\bf t})}\Bigr ),
\eeq
so we should consider
\beq\label{trig2}
\tau (x, {\bf t})=\prod_{i=1}^N \sigma (x-x_i)e^{\frac{1}{6}\, \gamma^2
(x-x_i)^2 +\gamma (x+x_i)}.
\eeq
Wit this choice, equation (\ref{e7}) acquires the form $k=z+\zeta (\lambda )+
\frac{1}{3}\, \gamma^2\lambda$ or
\beq\label{trig3}
k=z+\gamma \coth (\gamma \lambda ).
\eeq
The trigonometric limit of the function $\Phi (x, \lambda )$ is
$$
\Phi (x, \lambda )=\gamma \Bigl (\coth (\gamma x)+\coth (\gamma \lambda )\Bigr )
e^{-\gamma x \coth (\gamma \lambda )}.
$$
Therefore, the equation of the spectral curve can be written in the form
\beq\label{trig4}
\det \Bigl (W^{1/2}LW^{-1/2} +\gamma (1-\coth (\gamma \lambda ))(E-I)-
(z+\gamma \coth (\gamma \lambda ))I\Bigr )=0,
\eeq
where
$W=\mbox{diag}(w_1, w_2, \ldots , w_N)$ and
\beq\label{trig5}
L_{ij}=-\delta_{ij}p_i -\frac{(1-\delta_{ij})\gamma}{\sinh (\gamma (x_i-x_j))}=
-\delta_{ij}p_i -2\gamma (1-\delta_{ij})\, \frac{w_i^{1/2}w_j^{1/2}}{w_i-w_j}
\eeq
is the Lax matrix of the trigonometric Calogero-Moser model. Here and below we use the
notation $w_i=e^{2\gamma x_i}$.

After the transformations similar to the rational case equation (\ref{trig4})
can be brought to the form
$$
\gamma (1-\coth (\gamma \lambda ))\mbox{tr}\left [W^{-1/2}EW^{1/2}
\frac{1}{zI-(L-\gamma I)}\right ]=1
$$
or 
\beq\label{trig6}
\lambda =\frac{1}{2\gamma}\, \log \left [1-2\gamma
\mbox{tr}\left (W^{-1/2}EW^{1/2}
\frac{1}{zI-(L-\gamma I)}\right )\right ].
\eeq
Applying the formula $\det (I+Y)=1+\mbox{tr}\, Y$ for any matrix $Y$ of rank 1 in the
opposite direction, we have
\beq\label{trig7}
\lambda =\frac{1}{2\gamma}\, \log \det \left [I-2\gamma
W^{-1/2}EW^{1/2}
\frac{1}{zI-(L-\gamma I)}\right ].
\eeq
Now we are going to use the identity
\beq\label{trig8}
[L,W]=2\gamma (W^{1/2}EW^{1/2}-W)
\eeq
which can be easily checked. With the help of this identity, we can transform (\ref{trig7})
as follows:
$$
2\gamma \lambda =\log \det \left (
I-W^{-1}LW \frac{1}{zI-(L-\gamma I)}+\frac{L}{zI-(L-\gamma I)}-
\frac{2\gamma}{zI-(L-\gamma I)}\right )
$$
$$
=\log \det \left [ \left (I-\frac{2\gamma}{zI-(L-\gamma I)}\right )
\frac{zI-(L-\gamma I)}{zI-(L+\gamma I)}\right.
$$
$$
\left. \times \left (
I-W^{-1}LW \frac{1}{zI-(L-\gamma I)}+\frac{L}{zI-(L-\gamma I)}-
\frac{2\gamma}{zI-(L-\gamma I)}\right )\right ]
$$
$$
=\log \det \left [ \left (I-\frac{2\gamma}{zI-(L-\gamma I)}\right )
\left (I-W^{-1}LW \frac{1}{zI-(L+\gamma I)}+\frac{L}{zI-(L+\gamma I)}\right )\right ]
$$
$$
=\log \det \left [ \left (I-\frac{2\gamma}{zI-(L-\gamma I)}\right )
\frac{1}{zI-(L+\gamma I)}\Bigl (zI-(L+\gamma I)-W^{-1}LW +L\Bigr )\right ]
$$
$$
=\log \det \frac{zI-(L+\gamma I)}{zI-(L-\gamma I)}.
$$
Therefore, we get
\beq\label{trig9}
\begin{array}{c}
\displaystyle{\lambda =\frac{1}{2\gamma}\, \mbox{tr} \Bigl (\log (I-z^{-1}(L+\gamma I))-
\log (I-z^{-1}(L-\gamma I))\Bigr )}
\\ \\
\displaystyle{=-\frac{1}{2\gamma}\, \mbox{tr} \sum_{m\geq 1}\frac{z^{-m}}{m}
\Bigl ((L+\gamma I)^m -(L-\gamma I)^m \Bigr )}
\end{array}
\eeq
and
\beq\label{trig10}
H_m=\frac{1}{2\gamma (m+1)}\, \mbox{tr} \Bigl ((L+\gamma I)^{m+1} -(L-\gamma I)^{m-1} \Bigr )
\eeq
which agrees with the result of paper \cite{Z19a}.

\section*{Acknowledgments}

\addcontentsline{toc}{section}{\hspace{6mm}Acknowledgments}

We thank I. Krichever for illuminating discussions.
The research of A.Z. has been funded within the framework of the
HSE University Basic Research Program and the Russian Academic Excellence Project '5-100'.

\end{document}